\documentstyle[preprint,aps]{revtex}

\begin{document}
\draft
\preprint{SU-TS-9502}
\title{Insulator-to-metal transition in Kondo insulators under strong
magnetic field
\footnote{To appear in Phys. Rev. B}
}
\author{Tetsuro Saso\cite{Saso} and Masatoshi Itoh\cite{Itoh}}
\address{
Department of Physics, Faculty of Sciences, Saitama University, Urawa,
338 Japan }
\date{September 5, 1995}

\maketitle

\begin{abstract}
Magnetization curve and changes of the single-particle excitation
spectra by magnetic field are calculated for the periodic Anderson
model at half-filling in infinite spatial dimension by using the exact
diagonalization method. It is found that the field-induced
insulator-to-metal transition occurs at a critical field $H_c$, which
is of the order of the single ion Kondo temperature. The transition is
of first order, but could be of second order in the infinite system
size limit. These results are compared with the experiments on the
Kondo insulator YbB$_{12}$.
\\
\end{abstract}


\narrowtext
Study of the strongly correlated electron systems is one of the main
issues in condensed matter physics. The
heavy fermion materials\cite{heavy} are the typical examples of such systems,
in which the conduction electrons mix with the almost
localized 4f or 5f electrons, and form the strongly renormalized
quasi-particles which have effective mass of 100 to 1000 times larger
than the bare value. This strong renormalization is mainly due to the
local Kondo-type processes, which should be suppressed if an energy
gap opens at the Fermi level.

Recently, the materials called Kondo insulators are attracting much
interest. SmS\cite{SmS}, SmB$_6$\cite{SmB6}, YbB$_{12}$\cite{YbB12},
TmSe\cite{TmSe}, Ce$_3$Bi$_4$Pt$_3$\cite{Ce3Bi4Pt3},
CeNiSn\cite{CeNiSn}, etc. belong to this category.  These materials
are considered to be the intermediate-valence systems\cite{heavy}, but
show the activation type increase of the resistivity at low
temperatures, leading to insulating ground states, instead of the
metallic behaviors in the ordinary heavy fermion or
valence-fluctuating materials. In these Kondo insulators, a gap opens
at Fermi level because of the mixing of the conduction electrons with
the periodic assembly of the localized electrons. The single-site
Kondo processes, however, are still possible if the gap is comparable
to or smaller than the Kondo temperature\cite{Ogura93}. In addition,
the Kondo processes reduce the mixing, so that the gap is renormalized
to the value of the order of the Kondo temperature. Thus the Kondo
effect and the renormalization of the gap take place in a
self-consistent way.  It is observed that the specific heat
coefficient $\gamma$ (extrapolated from the temperatures higher than
the gap) is enhanced to some extent in the Kondo
insulators\cite{CeNiSn}, which indicates an importance of the
Kondo-type renormalization in these systems.

To further investigate the characters of the gap, an
interesting experiment was done on YbB$_{12}$\cite{Sugiyama88}, in
which the application of the strong magnetic field of up to 55T
destroys the gap and the system goes a transition from insulator to
metal. The magnetization starts to steeply increase at the critical
field. These experiments stimulate us to understand how the
above-mentioned states with strong renormalization will be modified or
destroyed by the magnetic field from the theoretical point of view.
For these purposes, we utilize the recently developed ideas and
techniques to treat the strongly correlated electron systems in the
infinite spatial dimensions\cite{Metzner89}.  We have calculated the
single-particle excitation spectra to determine the energy gap, which
vanishes at certain critical field $H_c$, indicating the
insulator-to-metal transition, and the magnetization curves, which
exhibits a steep rise at $H_c$. These results are consistent with the
experiments on YbB$_{12}$. Furtheremore, we have showed that the Van
Vleck susceptibility is finite even in insulators, and is enhanced by
the electron correlation.

We start with the periodic Anderson model (PAM),
\begin{eqnarray}
  H &=& \sum_{ij\sigma} t_{ij} c_{i\sigma}^+c_{j\sigma} + \sum_{i
\sigma} E_{f\sigma} f_{i \sigma}^+ f_{i \sigma} \nonumber \\
  &+& \frac{V}{\sqrt{N}} \sum_{i\sigma} \Bigl( f_{i\sigma}^+
c_{i\sigma} + h.c. \Bigr)
  + U \sum_i n_{f i \uparrow} n_{f i \downarrow}, \label{eq:pamha}
\end{eqnarray}
where $E_{f\sigma}=E_f-\sigma h$, $h=\mu_B H$ and $H$ denotes the
magnetic field.    The other notations are standard.  We neglect the
orbital degeneracy and assume the symmetric case $E_f=-U/2$ together
with half-filling condition to express the simplest model to the Kondo
insulators.  Note that we apply magnetic field only to f-electrons.
The reason for this is that the g factor for f-electrons, $g_f$,
differ from that for conduction electrons, $g_c$, so that the total
magnetization is not a conserved quantity.  The total susceptibility
is given by $\chi=[(g_f-g_c)/g_f]\chi_f$, where $\chi_f$ is the
susceptibility in which the magnetic field is applied only to
f-electrons\cite{Hirashima94}. If we take $g_f=g_c$ and apply magnetic
field to both electrons, $\chi$ vanishes for the half-filling at $T
\rightarrow 0$, which apparently contradicts with experiments.
We therefore calculate only $\chi_f$ and denote it simply as $\chi$ in
the following. In more realistic models, one has to take account of
the real values of g factors for both electrons, and also the values
of the magnetic moments in the crystal field split
states\cite{Nakano91}.

The first term in eq.(\ref{eq:pamha}) denotes
the kinetic energy of conduction electrons and is scaled in such a way
that in $d
\rightarrow \infty$ limit ($d$ denotes the spatial dimensions) the
density of states (DOS) has the form  $\rho(\epsilon)=(2/\pi
W)\sqrt{1-(\epsilon/W)^2}$, where $W=2\sqrt{d}t$, if the Bethe
lattice with the hopping $t$ parameter is assumed.   We, however,
regard it as a model DOS for the real system.  Furthermore, this shape
is more suitable for the discussion of the insulator-metal transition,
since it yields the sharp transition in contrast to the Gaussian DOS.
We also assume in the following that any magnetic order is suppressed
because of an insufficient nesting or strong frustrations in real
systems, so that we study only the paramagnetic phase. $W=\hbar=k_B=1$
is assumed throughout the present paper.

We replace $E_f$ by $\tilde{E}_f=E_f+U/2=0$ in the Hamiltonian,
assuming the electron-hole symmetry, and add the term
$(U/2)\sum_{i\sigma} n_{f i \sigma}$ to compensate the above
modification.  But the latter term is constant because of the
electron-hole symmetry even under finite field since $n_{f i
\sigma}=(n_f+\sigma m)/2$ and $n_f=1$,
where $m=n_{f i \uparrow}-n_{f i \downarrow}$ defines the
magnetization. The thermal Green's function for f-electrons is
expressed as
\begin{equation}
  G_{f\sigma}(i\epsilon_n) = \int d\epsilon_k
 \frac{\rho(\epsilon_k)}{G_{f\sigma}^0(k,i\epsilon_n)^{-1}
 -\Sigma_\sigma(i\epsilon_n)},
\end{equation}
where $G_{f\sigma}^0(k,i\epsilon_n) = (i\epsilon_n-\tilde{E}_f+\sigma h
- V^2/(i\epsilon_n-\epsilon_k))^{-1}$ denotes the unperturbed
f-Green's function. In $d \rightarrow \infty$, the lattice problem is
reduced to solving a generalized Anderson impurity embedded in an
effective medium self-consistently\cite{Jarrell92,Rozenberg92}. The
unperturbed Green's function for this impurity is given by
$\tilde{G}_{f\sigma}(i\epsilon_n)^{-1}
=G_{f\sigma}(i\epsilon_n)^{-1}+\Sigma_\sigma(i\epsilon_n)$. We can
fit $\tilde{G}_{f\sigma}(i\epsilon_n)^{-1}$ by that of the impurity
Anderson model with finite number of conduction electron levels
$\epsilon_\ell$ and energy-dependent mixing $V_\ell$ ($\ell=1\sim
N_s$): $\tilde{G}_{f\sigma}^{fit}(i\epsilon_n)^{-1} = (i\epsilon_n
-E_\sigma +\sigma h -
\sum_{\ell\sigma}v_{\ell\sigma}^2/(i\epsilon_n
-\epsilon_{\ell\sigma}))$\cite{Caffarel94}.
Accuracy of this fitting is generally very well in spite of
the small number of levels. It is because not only the mixing
$V_\ell$'s but also the energy levels $\epsilon_\ell$'s are treated as
the fitting parameters.  As a result, bulk quantities can be obtained
with good accuracy in the present method, although discreteness of the
levels reflects in dynamical quantities. $E_\sigma$ above denotes the
effective f level, but the fitting always yields $E_\sigma=0$.
Although their method is applicable also to the finite temperatures,
we investigate only the zero temperature properties.
The ground state of this Anderson model was determined by the modified
Lanczos method\cite{Dagotto85} and the thermal Green's function
${G}_{f\sigma}(i\epsilon_n)$ was calculated in the form of the
continued fraction by the Lanczos method\cite{Gagliano87}.  It is
known that $N_s=7$ is sufficient to obtain a good fit to the thermal
Green's function by that with finite levels for the Hubbard
model\cite{Caffarel94} and the same holds true in the present case.  A
new estimate of the self-energy is obtained by
$\Sigma_\sigma(i\epsilon_n) =\tilde{G}_{f\sigma}(i\epsilon_n)^{-1} -
G_{f\sigma}(i\epsilon_n)^{-1}$, which completes the self-consistent
loop.

We first present the single-particle excitation spectrum for $U=2$ and
$h=0$ in Fig.1(a). $V=0.5$ is used in the following calculations. It
should be noted that the bare f-electron peaks at $\pm U/2$ are
shifted to deeper positions, in accord with the quantum Monte Carlo
calculation\cite{Jarrell93a}, the numerical renormalization group
calculation\cite{Sakai95} and the self-consistent second order
perturbation theory\cite{Hirashima94}. In addition to them, a Kondo
resonance peak is formed around $\epsilon=0$, which is split by the
mixing with conduction electrons. The size of the gap, $E_g$ are
plotted in Fig.2 as a function of $U$. $E_g$ is about twice the Kondo
temperature $T_K$ for the single impurity for large $U$.  Here we
define $T_K$ by $T_K=1/4\chi$, where $\chi$ is the magnetic
susceptibility obtained by Bethe Ansatz for the impurity Anderson
model\cite{Horvatic85}.

It should be noted that the self-energy $\Sigma_\sigma(i\epsilon_n)$
is proportional to $-i\epsilon_n$ even in the insulating phase. This
is in contrast to the case of the Hubbard model, in which
$\Sigma_\sigma(i\epsilon_n) \propto -i\epsilon_n$ in the metallic
phase but $\propto -i/\epsilon_n$ in the insulating
phase\cite{Rozenberg92}. A reason to this difference would be that the
gap is formed by the mixing in PAM and it exists even for $U=0$
whereas it is purely due to the correlation in the Hubbard model.

Therefore, the above spectra are well represented by the renormalized
mean-field (MF) theory (eq.(\ref{rho_f}) below with $h=0$), in which
$V^2$ in the mixed band dispersion
\begin{equation}
  E^{(\pm)}(\epsilon_k)=\frac{1}{2}\left[ \epsilon_k+\tilde{E}_f \pm
\sqrt{(\epsilon_k-\tilde{E}_f)^2+4V^2}\right],
\end{equation}
is replaced with $zV^2$, where $z$ is the renormalization factor,
$z^{-1}=1-\partial
\Sigma_\sigma(i\epsilon_n)/\partial i\epsilon_n|_{\epsilon_n\rightarrow 0}$.
For $U=2$ we found $z=0.57$. The energy
gap
is given by
$E_g=E^{(+)}(\epsilon_k=-W)-E^{(-)}(\epsilon_k=+W)=-W+\sqrt{W^2+4zV^2}$
for the symmetric case. It reads $E_g \approx 2zV^2/W$ when $V \ll W$.

Next, we apply magnetic field. Fig. 2 displays the magnetization curve
of the f-moment $m=n_{f\uparrow}-n_{f\downarrow}$ for U=0, 1, 2 and 3.
The dotted curve is the result for the infinite system with $U=0$
obtained from
\begin{equation}
  m = \int d\epsilon f(\epsilon)
[\rho_{f\uparrow}(\epsilon)-\rho_{f\downarrow}(\epsilon)],
\label{eq:m}
\end{equation}
where $f(\epsilon)$ is the Fermi function and
$\rho_{f\sigma}(\epsilon)$ denotes the f-component density of states
of $\sigma$ spin electrons under magnetic field,
\begin{equation}
  \rho_{f\sigma}(\epsilon)=\frac{1 \mp b_\sigma}{1 \pm b_\sigma}
\rho(E_\sigma(\epsilon)) \label{rho_f}
\end{equation}
(upper (lower) sign for $\epsilon>\tilde{E}_{f\sigma}$
($\epsilon<\tilde{E}_{f\sigma}$)),
\begin{equation}
  b_\sigma = (E_\sigma(\epsilon)-\tilde{E}_{f\sigma}) /
  \sqrt{(E_\sigma(\epsilon)-\tilde{E}_{f\sigma})^2+4V^2},
\end{equation}
$E_\sigma(\epsilon)=\epsilon-V^2/(\epsilon-\tilde{E}_{f\sigma})$ and
$\tilde{E}_{f\sigma}=\tilde{E}_f-\sigma h$. This formula gives the f
component susceptibility $\chi_f=0.76676$, which agrees with the
initial slope of the curve for $U=0$ in Fig. 3.

For $U=0$, the spectrum consists of only a few sharp peaks due to the
smallness of the system, so that the magnetization curve shows
redundant steps, although its gross feature agrees well with
eq.(\ref{eq:m}). For finite $U$, however, the spectra acquire finer
structures and $m(h)$ becomes smoother, although a redundant two-step
jump is still seen at $U=1$.  Beside these artifacts, the
magnetization curves indicate sharp increases at the critical field
$h_c$, which is plotted in Fig. 2 as a function of $U$.  The change of
the spectra across $h_c$ are displayed in Fig. 1(b) ($h {<\atop\sim}
h_c$) and (c) ($h {>\atop\sim} h_c$), indicating that the up- and the
down- spin bands shift with each other and the gap closes. Thus the
system undergoes a transition from insulator to metal at $h_c$.

We also show in Fig. 4 how the gap $E_g(h)$ closes by magnetic field.
It is seen that the gap closes as $E_g(h)=E_g(0)-2h$, but for $U>0$
(and maybe also for $U=0$ around $h=0.26$), $E_g$ jumps to zero before
$E_g(0)-2h$ vanishes, indicating a first order transition. By a naive
argument we expect that the Kondo renormalization would be suppressed
by the magnetic field, so that $z$ and the effective mixing increase,
hence the gap increases, whereas the up- and down-spin bands are
relatively shifted by the field, so that the gap reduces.  Hence the
critical field $h_c$ is determined as a result of the competition of
these two effects. The present calculation shows that the gap
decreases in proportion to $-2h$, which indicates that $z$ does not
change but the gap is closed because of the rigid shift of the each
spin band.  In fact, we found that the self-energy scarcely changes
its shape for $h<h_c$.

Whether the transition is of 1st order is a delicate issue.
Investigating the details of the spectra shown in Figs.1(b) and (c),
we found that a small peak discontinuously moves across the chemical
potential, $\epsilon=0$, when $h$ goes through $h_c$. Since we have
electron-hole symmetry and even number of electrons in total, no state
can appear precisely at $\epsilon=0$. Therefore, in the finite system
calculation, a continuous change of $h$ across $h_c$ may cause a
discontinuous shift of peaks across $\epsilon=0$, yielding a
discontinuous closing of a gap. Thus, we expect that the transition
could become second order in the infinite size limit, but we leave it
open for further study.

Sugiyama, et al.\cite{Sugiyama88} analysed their experiment on
YbB$_{12}$ and concluded that the gap decreases slightly faster than
linearly with the field (see the insert in Fig.4). Since there are two
activation energies in their resistivity data, they introduced a fine
structure in the model density of states characterized by two gap
values $2\Delta_1=60$ K and $2\Delta_2=180$ K. The average of the two
gives $E_g \sim 120 K$ and the critical field $H_c \sim 450$ KOe
corresponds to 57 K, using $gJ=1.9$. Thus they satisfy $E_g \sim
2H_c$.  Our calculations are in good agreement to these features at
least qualitatively.

Finally, we show in Fig.5 the f-component magnetic susceptibility
obtained from the initial slopes of Fig.3, together with that
multiplied by $E_g$ and $z$. Note that the susceptibility has finite
values despite a finite gap opens, since the total magnetization is
not conserved in the present model. It is sometimes called the Van
Vleck susceptibility. Kontani and Yamada\cite{Kontani95} gave a
discussion that it should be enhanced approximately by $z^{-1}$. Our
result shows that it is indeed enhanced  by the interaction, although
the enhancement becomes weaker than $z^{-1}$ or $E_g^{-1}$ for strong
coupling regime, which may be due to the interspin exchange
interaction. The present result is qualitatively in agreement with the
self-consistent perturbational result\cite{Hirashima94}.

In summary, we have calculated the magnetization curve and the
single-particle excitation spectra of the periodic Anderson model at
half-filling in infinite dimensions in the paramagnetic phase, which
is a prototype of the Kondo insulators, and found that the
insulator-to-metal transition takes place by applying magnetic field.
The gap decreases as $E_g(h)=E_g(0)-2h$, and vanishes at the critical
field $h_c$, which is found to be of the order of $T_K/2$. It
indicates that the gap closes for $h<h_c$ in such a way that each spin
band is shifted rigidly by $\pm h$. The transition is of first order,
but could be of second order in the infinite system. The above
calculation are mostly in accordance with the observations in
YbB$_{12}$,
although more realistic model must be used for detailed comparison.
Further study of the magnetic field effects to the states
away from half-filling will also shed light on the metamagnetism in
metallic heavy fermion systems\cite{Mignot88}.

This work is supported by Grant-in-Aid for Scientific Research on
Priority Areas, No. 07233209, ``Physics of Strongly Correlated
Metallic Systems", from the Ministry of Education, Science and
Culture.
The computation was done using FACOM VPP500 in the Supercomputer Center,
Institute for Solid State Physics, University of Tokyo.
One of the authors (T. S.) thank Professor O. Sakai, Dr. K. Sugiyama
and Mr. T. Mutou for useful conversations.

\newpage
\begin{figure}
\caption{The single-particle spectrum of PAM for $U=2$, $V=0.5$ and
(a) $h=0$, (b) $h=0.09$ and (c) $h=0.1$.
The dotted line in (a) indicates the spectrum of the mean-field theory
with reduced mixing $zV^2$ with$z=0.57$.}
\label{fig1}
\end{figure}

\begin{figure}
\caption{The energy gap $E_g$, the twice of the critical magnetic field
$2h_c$ and the impurity Kondo temperature $T_K$ by the Bethe Ansatz are
plotted against $U$.}
\label{fig2}
\end{figure}

\begin{figure}
\caption{The magnetization curve for $U=0$(circle), 1(triangle),
2(square) and 3(diamond).}
\label{fig3}
\end{figure}

\begin{figure}
\caption{The energy gap for $U=0$(circle), $U=1$(triangle), 2(square)
and 3(diamond) are plotted as a function of the magnetic field.
The dotted lines
indicate $E_g(h=0)-2h$. The insert is the gap $2\Delta_1$ obtained by
Sugiyama, et al. for YbB$_{12}$.}
\label{fig4}
\end{figure}

\begin{figure}
\caption{The f-component magnetic susceptibility (filled circles)
and that multiplied by $E_g$ (open circles) and by $z$ (square) scaled
at $U=0$.}
\label{fig5}
\end{figure}

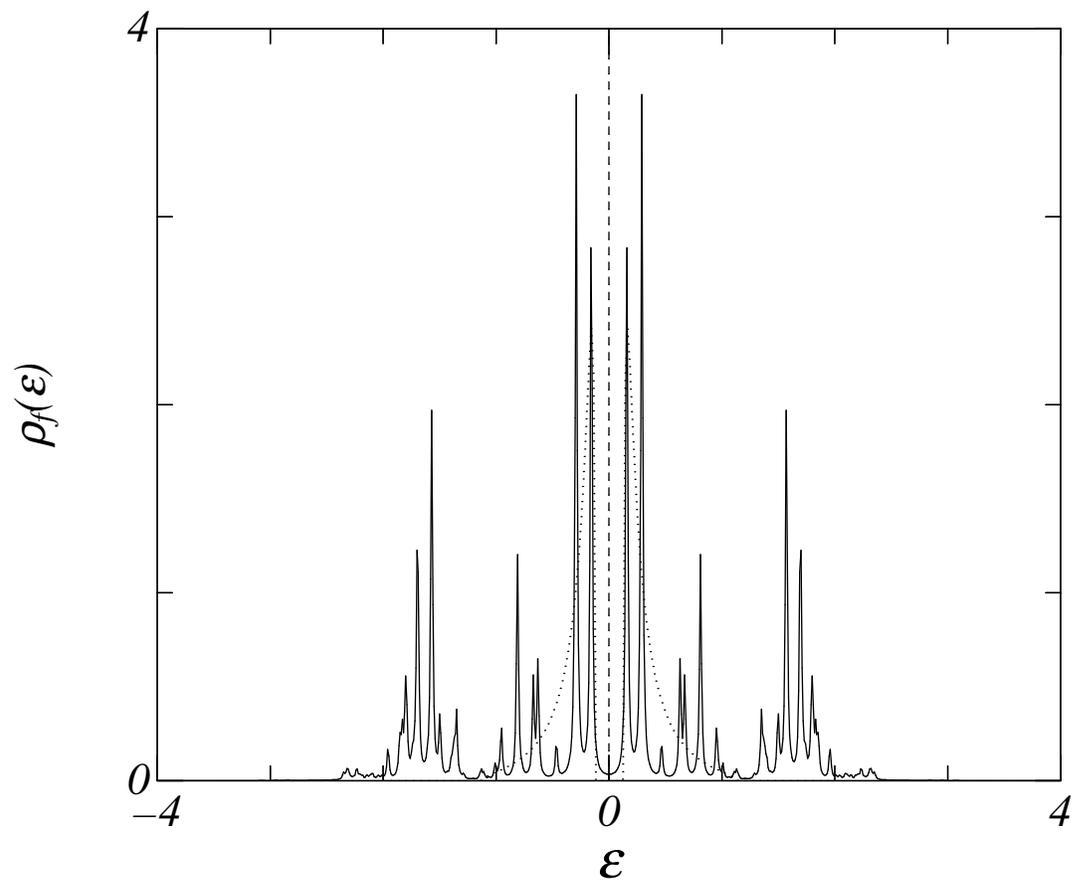
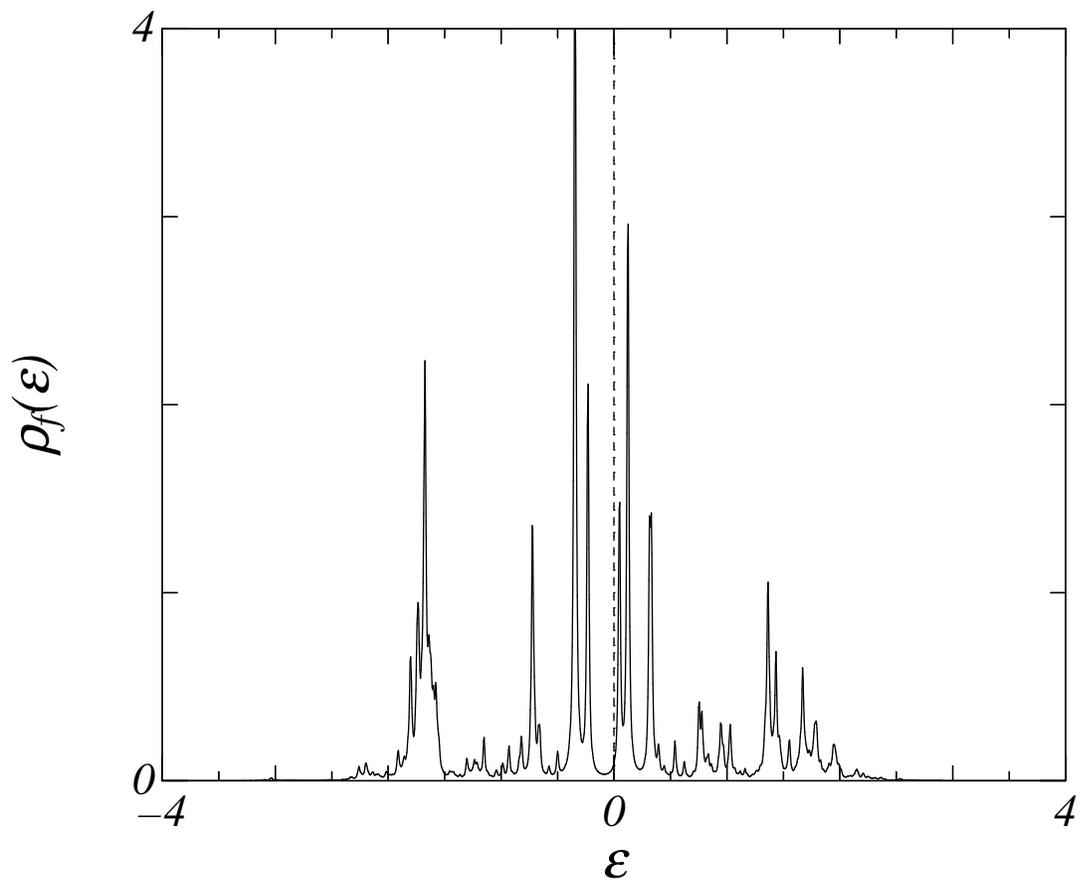
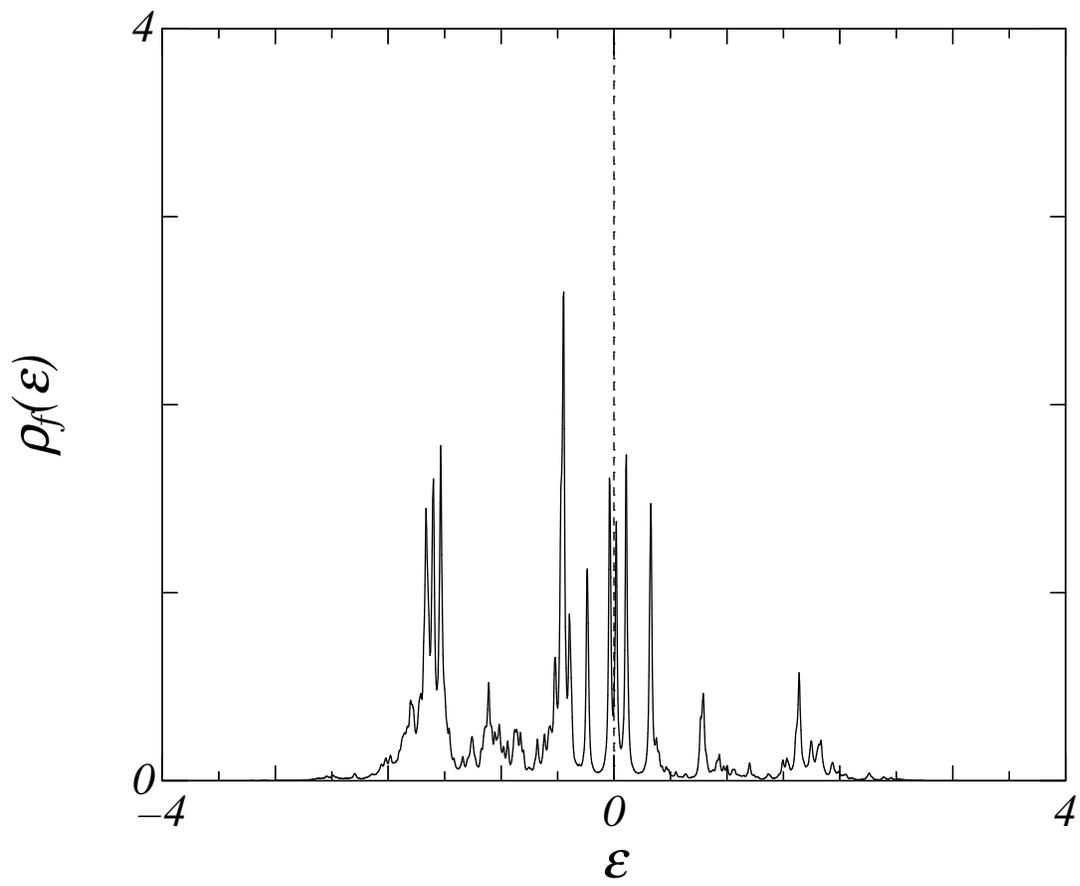
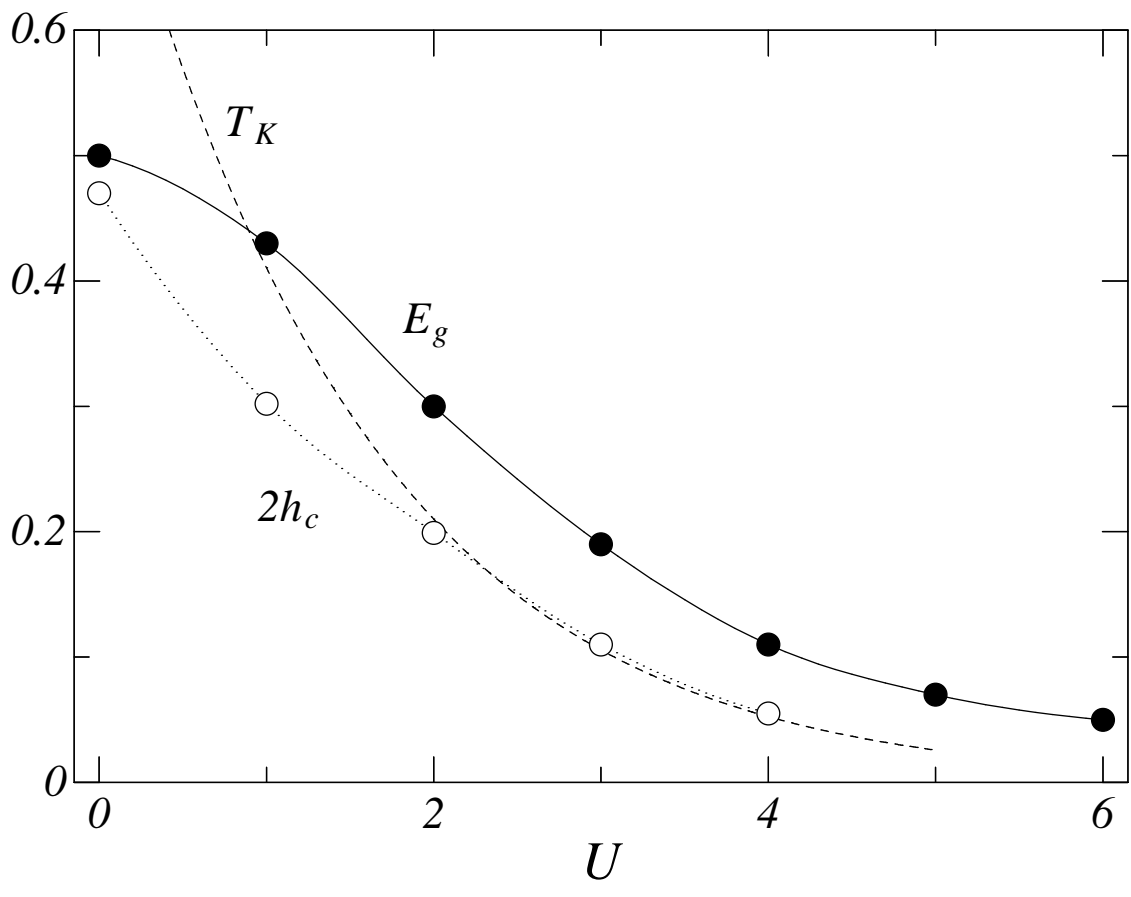
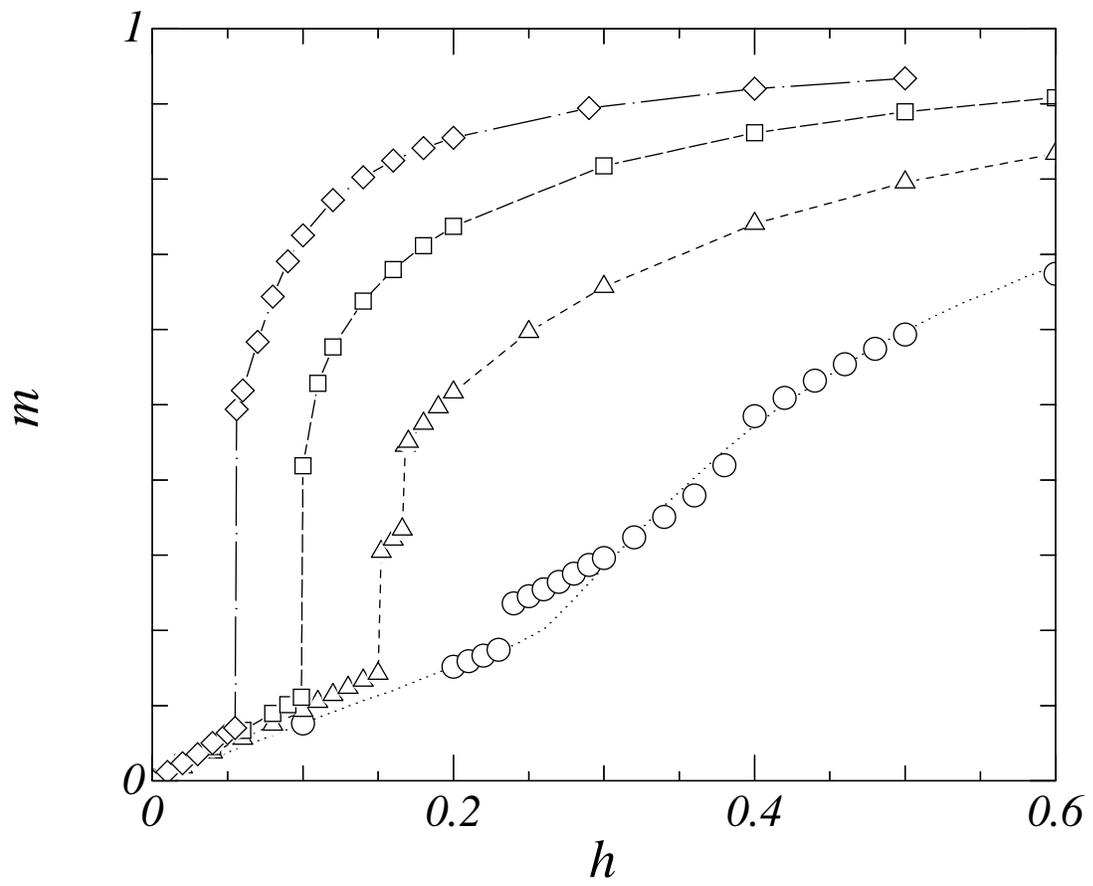
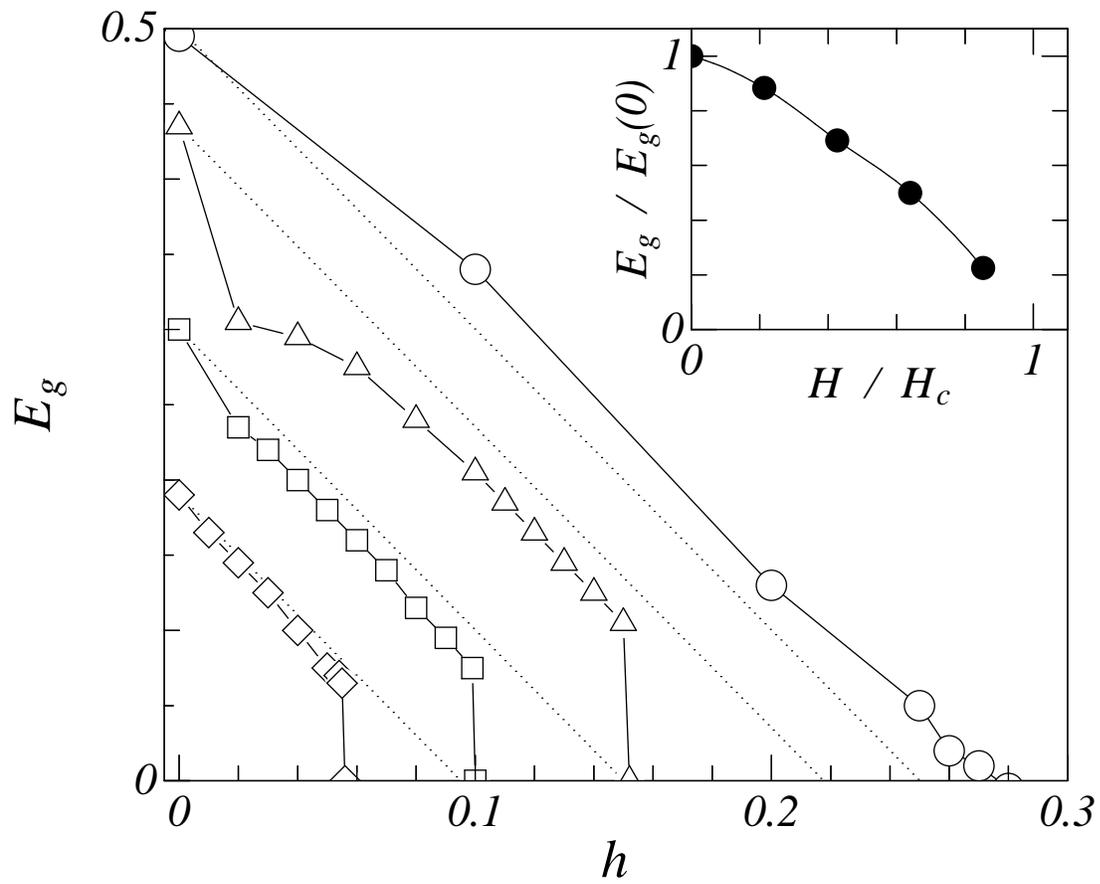
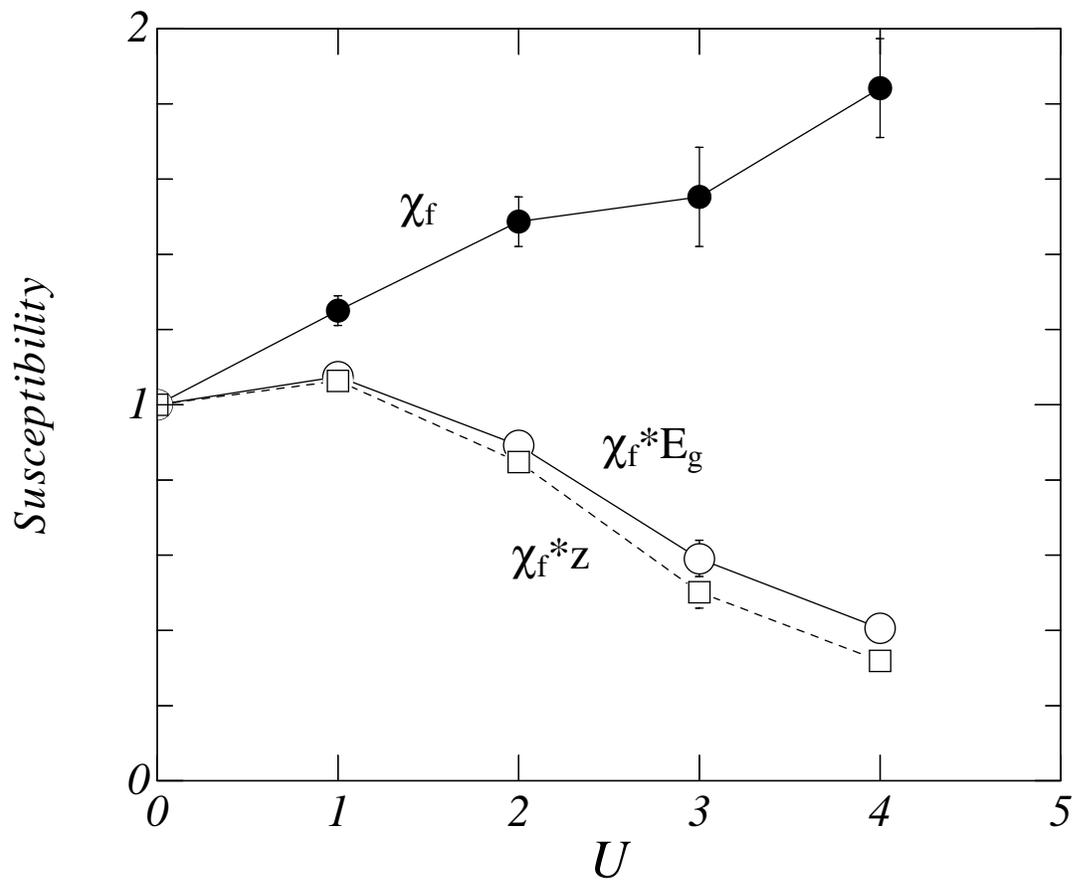
\end{document}